# A molecular dynamics simulation study of thermal conductivity of plumbene

Rafat Mohammadi[*a], Behrad Karimi[a], John Kieffer[b], Daniel Hashemi[c]

[a] Department of Mechanical Engineering, Faculty of Engineering, Arak University, Arak 38156-88349, Iran

[b] Department of Materials Science and Engineering, University of Michigan, Ann Arbor, MI, 48109, USA

[c] Department of Physics and Optical Engineering, Rose-Hulman Institute of Technology, IN 47803, USA

**Abstract**

We investigate the thermal conductivity of plumbene using molecular dynamics simulations, overcoming existing limitations by optimizing the parameters of Tersoff and Stillinger-Weber potentials via artificial neural networks. Our findings indicate that at room temperature, the thermal conductivity of a 1050 Å × 300 Å plumbene sheet is approximately 8 W/m.K , significantly lower (23%) than that of bulk lead. Our analysis elucidates that thermal conductivity is enhanced by increased sample length, while it is reduced by temperature. Moreover, plumbene samples with zigzag edges display superior thermal conductivity compared to those with armchair edges. In addition, the thermal conductivity of plumbene exhibits an increase at low tensile strains, whereas it decreases as the strains become larger. This investigation provides crucial insights into the thermal conductivity behavior of plumbene under varying conditions.

## 1. Introduction

Materials scientists are tirelessly working towards the development of advanced materials to meet the demands of future technological applications. Among these materials, two-dimensional (2D) materials, also referred to as monolayers or single-layer materials, exhibit exceptional physical properties that set them apart from their bulk counterparts. Graphene, the pioneering 2D material, stands out for its remarkable stability, conductivity, and flexibility [1]. Plumbene, a lead-based counterpart to graphene, has recently emerged as a subject of intense interest due to its unique structure and potential for electronic applications [2-9].

Both graphene and plumbene possess a hexagonally arranged structure, but plumbene has a slightly buckled configuration, leading to increased overlap between its sigma (r) and pi (p) orbitals. Plumbene presents three stable structures: the planar form, as well as the high-buckled (HB) and low-buckled (LB) honeycomb structures. Notably, the buckled configurations exhibit greater stability compared to the planar structure. The low-buckled plumbene has a lattice constant of 4.934 Å and a buckling height of 0.99 Å, whereas the high-buckled plumbene measures 3.692 Å and 2.55 Å for its lattice constant and buckling height, respectively [3,6]. In a noteworthy experiment conducted by Yuhara et al. in 2019, low-buckled honeycomb structures of plumbene were initially synthesized [7].

Plumbene holds immense promise as a versatile platform for fabricating novel topological insulators and large-gap quantum spin Hall (QSH) insulators [2-6]. Furthermore, it has emerged as a potential candidate for hydrogen storage [9]. With its distinct properties and growing scientific interest, plumbene represents an exciting avenue for exploring cutting-edge technological advancements. However, despite its potential, little is known about the different properties of plumbene. Lu et al. [3] used *Ab initio* density functional theory (DFT) calculations to investigate the material's electronic and structural properties of functionalized atomic lead films. They demonstrated that the 2D Pb structure can have applications in topological electronic devices. Li et al. [6] investigated electronic and topological properties of plumbene by combining tight-binding models with first-principles calculations. Constructive coupling effects of topological states found in the low-buckled plumbene, causing the system to be a normal insulator, opposite to topologically nontrivial states formed in other 2D group IVA monolayers (from graphene to stanene). Bechstedt et al. [10] used a four-band model to investigate the thermal and electrical conductance of 2D electron gases in doped Xenes, including graphene, silicene, germanene, stanene, and plumbene. They showed that thermal conductance increases linearly or quadratically with temperature. Mahdavifar et al. [11] used first-principles calculations to examine the electronic and mechanical properties of single-layer plumbene in different armchairs, zigzag, and biaxial directions under applying external strain and electric field and in the presence of a Stone-Wales defect. They found that the plumbene monolayer exhibits the lowest Young's modulus and ideal strength among group IVA mono-elemental monolayers. This is attributed to its relatively weaker bond length compared to its counterparts. They also suggested that plumbene can be used in flexible electronics and is a promising candidate for the quantum spin hall effect when its electronic properties are tuned using external fields. Zhang et al. [12] calculated the superconducting critical temperature of HB plumbene using first-principles calculations. They showed that the critical temperature of HB plumbene can be tuned by adjustment of Fermi level.

Complementing the aforementioned first-principle calculations and experimental/theoretical research on plumbene, two distinct studies have leveraged molecular dynamics (MD) simulations to elucidate the properties of plumbene. Das et al. [13] investigated the mechanical properties of plumbene, including yield strength, ultimate tensile strength, and Young's modulus using MD simulations and studied the effects of sample size, temperature, and strain rate on these properties. Das et al. [14] also carried out MD simulations to predict the thermal properties of plumbene and studied the effects of sample size on thermal conductivity. This group examined the thermal conductivity of plumbene for samples with widths of 300 Å and lengths ranging from 600 Å to 1050 Å. Their results showed that plumbene thermal conductivity fluctuates between 0.04 W/m.K and 0.24 W/m.K for different sample lengths. The embedded atom method (EAM) potential with the parameters of the bulk lead was used in these two MD simulations of plumbene.

Tersoff and Stillinger-Weber (SW) potentials are the most common potentials used in previous MD simulations on other 2D materials of group IVA [15-20]. However, the parameters of these potentials have not yet been determined for lead or plumbene. Therefore, Das et al. used the available EAM potential of bulk lead for MD simulation of plumbene and no details about the structural parameters such as the lattice constant, bond length, and buckling height of the simulated plumbene have been mentioned in their research [13,14]. In addition, previous experimental observations [15] and MD simulations [15-20] of the thermal conductivity of other 2D materials of group IVA have shown that the thermal conductivity of the 2D materials of group IVA increases with increased length. While the simulated thermal conductivity of plumbene fluctuates with increased length in Das et al. research [14].

In this study, we tackle the challenges associated with plumbene MD simulations by optimizing the necessary parameters of Tersoff and SW potentials using artificial neural networks (ANNs). To validate these optimized potentials, we compare their obtained results against *Ab initio* molecular dynamics simulations conducted at different temperatures. Subsequently, employing non-equilibrium molecular dynamics (NEMD) simulations with the optimized potentials, we investigate the thermal conductivity of plumbene sheets with both armchair and zigzag configurations. Our simulations explore the impact of sample length, width, temperature, and tensile strain on plumbene's thermal conductivity.

It is worth noting that previous theoretical predictions in Refs [3,4,6,11] have classified low-buckled plumbene as a topologically trivial insulator. Considering the energetic stability and electronic properties, our study primarily focuses on analyzing the thermal conductivity of the low-buckled plumbene.

Plumbene's distinctive structure and spin-orbit interaction position it as a promising candidate for various electronic applications. However, to fully comprehend its thermal and mechanical properties, further research is required. MD simulations prove to be a valuable tool in this endeavor. Hence, our study aims to contribute to the understanding of plumbene's thermal conductivity under different conditions.

## 2. Simulation methods

In this study, various techniques were employed to investigate the thermal conductivity of plumbene. First, *Ab initio* molecular dynamics simulations were carried out to reproduce the buckling structure of plumbene. The necessary parameters of the SW and Tersoff potentials for plumbene were then optimized using an ANN. Finally, NEMD simulations were conducted to calculate the thermal conductivity of plumbene using the optimized SW and Tersoff potentials. The details of each of these methods are discussed in further detail in the following sections.

### 2.1. Potential Optimization

The Tersoff and SW potentials are widely recognized empirical approaches extensively employed in the field of MD simulations. These potentials have been developed based on many-body interactions and have been carefully calibrated using a variety of data sources, including *Ab initio* calculations, DFT, artificial intelligence, and experimental results. In the realm of MD simulations involving group IVA 2D materials, these potentials are among the most frequently used potentials

[15-20]. However, these potentials have not been previously utilized or optimized specifically for lead or plumbene.

In this study, we focus on optimizing the parameters of the Tersoff and SW potentials specifically for plumbene. To achieve this, we begin by conducting *Ab initio* molecular dynamics simulations to accurately reproduce the buckling structure of plumbene. Subsequently, we employ an in-house code to train a multi-layered perceptron neural network with six inputs. The input data points encompass atomic number, bond length, buckling height, atom bond angle, band gap, and lattice constant length. To design the network, the Levenberg-Marquardt algorithm is employed [21], and a specific number of iterations are carried out to obtain optimal potential parameters. To enhance the network's performance, the input vector is normalized. The accuracy of the network is evaluated using the root-mean-square (RMSE) error as a metric.

By optimizing the Tersoff and SW potentials parameters specifically for plumbene and employing a comprehensive approach involving *Ab initio* simulations and neural network training, we aim to enhance the accuracy and reliability of MD simulations for this unique material system.

### 2.1.1. *Ab initio* molecular dynamics calculations

The Vienna *Ab initio* simulation package (VASP)'s [22,23] projected augmented wave (PAW) [24] formalism, based on *Ab initio* DFT, is used to calculate the geometry and structure of the plumbene. The Perdew-Burke-Ernzerhof generalized-gradient approximation (GGA-PBE) is employed to describe the exchange and correlation functional [25]. These functionals have been previously employed to assess the stability of doping transition metals in plumbene [26,27]. The plane-wave cutoff energy is set to be 600 eV and a vacuum space of 15 Å is set to avoid the interactions between the two adjacent slabs. The calculation is performed with a unit cell containing 32 Pb atoms. All the Pb atoms in the low-buckled plumbene are allowed to relax until the Hellmann-Feynman force on each atom is smaller than 0.001 eV/Å. The centered Monkhorst-Pack grids of $6 \times 6 \times 1$ are adopted.

Here, our *Ab initio* MD calculation shows that at room temperature the bond length is 2.99 Å, the lattice constant is 4.98 Å, and the buckling height is 1.06 Å, which are comparable with previous research [3,6].

### 2.1.2. Optimization of Tersoff potential parameters

According to the Tersoff equation, the total potential energy of the system, E, is made up of pairwise contributions from the attraction [$f_A(r_{ij})$] and repulsive [$f_C(r_{ij})$] interactions between each unique atomic pair, i-j, at a distance $r_{ij}$. The energy equation is written as [28,29]:

$$E = \frac{1}{2} \sum_{i} \sum_{j \neq i} f_c(r_{ij})[f_R(r_{ij}) + b_{ij} f_A(r_{ij})] \tag{1}$$

The function $f_c(r_{ij})$ is the cut-off function that limits the range of interaction to the nearest neighbors. It is defined as:

$$f_c(r) = \begin{cases} 1 & r < R - D \\ \frac{1}{2} - \frac{1}{2}\sin(\frac{\pi(r-R)}{2D}) & R - D < r \leq R + D \\ 0 & r \geq R + D \end{cases} \tag{2}$$

where R and D specify the cut-off distance of the potential. The repulsive and attractive contributions to the bond energy of i-j pair decay exponentially with the separation distance $r_{ij}$, which is written as:

$$f_R(r) = Ae^{-\lambda_1 r} \tag{3}$$

$$f_A(r) = -Be^{-\lambda_2 r} \tag{4}$$

The term $b_{ij}$ in Equation (1) describes the bond-order for a pair of atoms i-j, which describes the weakening of an i-j bond owing to the presence of the other bond i-k around atom i. This term explicitly accounts for angular interactions via Equations (5-7):

$$b_{ij} = (1 + \beta^n \zeta_{ij}^n)^{\frac{-1}{2n}} \tag{5}$$

$$\zeta_{ij}^n = \sum_{k \neq i,j} f_c(r_{ik}) g_{ik}(\theta_{ijk}) e^{[\lambda_3^m (r_{ij} - r_{ik})^m]} \tag{6}$$

$$g(\theta_{ijk}) = \gamma_{ijk}\left(1 + \frac{c^2}{d^2} - \frac{c^2}{d^2 + (\cos\theta_{ijk} - h)^2}\right) \tag{7}$$

The parameters B, n, $\lambda_1$, $\lambda_2$, $\gamma$, c, d, and h are adjustable, while m=3.0, $\lambda_3 = 0$ and $\gamma=1.0$ are kept constant. In all, the 11 independent parameters have been optimized. The calculated parameters by ANN are ($h.R.D.A.B.\lambda_1.\lambda_2.\beta.n.c.d$) and shown in Table 1.

Table 1: Optimized Tresoff potential parameters obtained from ANN

| | |
|---|---|
| $R$ (Å) | 3.70 |
| $B$ (eV) | $4.30 \times 10^2$ |
| $A$ (eV) | $1.8501 \times 10^3$ |
| $d$ | $1.62137 \times 10^1$ |
| $n$ | $7.2402 \times 10^{-1}$ |
| $\beta$ | $3.70 \times 10^{-7}$ |
| $h$ | $-3.1998 \times 10^{-1}$ |
| $c$ | $3.016429 \times 10^5$ |
| $D$ (Å) | $1.27016895 \times 10^{-1}$ |
| $\lambda_1$ (1/ Å) | 2.445200282 |
| $\lambda_2$ (1/ Å) | 1.70 |

## 2.1.3. Optimization of Stillinger–Weber potential parameters

In SW potential, the energy of the system, E, is calculated by the following equation [29,30].

$$E = \sum_i \sum_{j>i} \Phi_2(r_{ij}) + \sum_i \sum_{i \neq j} \sum_{k>j} \Phi_3(r_{ij}.r_{ik}.\theta_{ijk}) \qquad (8)$$

where $\Phi_2$ is a two-body term and $\Phi_3$ is a three-body term. In general, in the formula, on all neighbors of j and k, the i atom is at a cutoff distance:

$$\Phi_2(r_{ij}) = A_{ij}\epsilon_{ij}\left[B_{ij}\left(\frac{\sigma_{ij}}{r_{ij}}\right)^{P_{ij}} - \left(\frac{\sigma_{ij}}{r_{ij}}\right)^{q_{ij}}\right]e^{\left(\frac{\sigma_{ij}}{r_{ij}-a_{ij}\sigma_{ij}}\right)} \qquad (9)$$

$$\Phi_3(r_{ij}.r_{ik}.\theta_{ijk}) = \lambda_{ijk}\epsilon_{ijk}[\cos\theta_{ijk} - \cos\theta_{0ijk}]^2 e^{\left(\frac{\gamma_{ij}\sigma_{ij}}{r_{ij}-a_{ij}\sigma_{ij}}\right)}e^{\left(\frac{\gamma_{ik}\sigma_{ik}}{r_{ik}-a_{ik}\sigma_{ik}}\right)} \qquad (10)$$

As indicated in Equations (9) and (10); to use the SW potential, the parameters $\epsilon$, a, $\gamma$, A, $\sigma$, $\lambda$, $\cos\theta_0$, B, p and q could be calculated. These parameters are optimized using the neural network and their values are shown in Table 2.

Table 2: Optimized Stellinger-Weber potential parameters obtained from ANN

| | |
|---|---|
| q | 0.00 |
| $\epsilon$ (eV) | 1.95 |
| a | 1.80 |
| $\gamma$ | 1.20 |
| p | 4.00 |
| $\lambda$ | $3.10 \times 10^1$ |
| $\sigma$ (Å) | 2.551 |
| A | 7.149556277 |
| $\cos\theta_0$ | $-3.13992456 \times 10^{-1}$ |
| B | $6.122245584 \times 10^{-1}$ |

## 2.2. MD Simulations

The Large-scale Atomic/Molecular Massively Parallel Simulator (LAMMPS) [31] was used to perform molecular dynamics simulations, while Avogadro [32] was used for atom visualization. NEMD simulation was employed to simulate thermal conductivity of plumbene using the optimized SW and Tersoff potentials. A view of the plumbene atomic structure is shown in Figure 1(A). In this figure, the hexagonal structure of zigzag plumbene has been illustrated. Figure 1(B) depicts the top and side views of LB plumbene. The buckling height, h, is also shown in Figure 1(B).

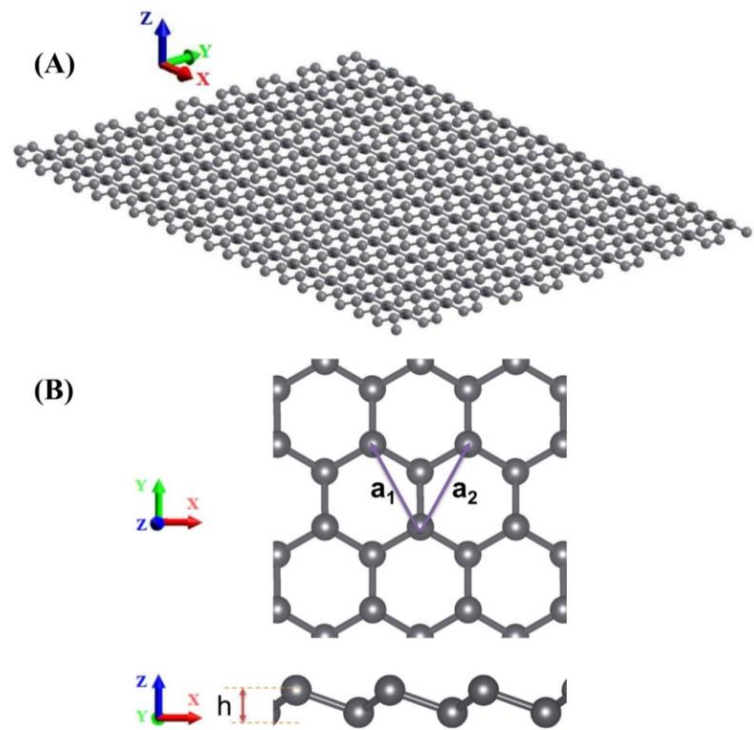

Figure 1: (A) Three-dimensional view of plumbene sheet. (B) The top and side views of buckled plumbene.

Figure 2 displays the simulation model for the thermal conductivity calculations. The simulation domain includes three regions, which are fixed walls, reservoir regions and the normal simulation region. At the two ends of the model, fixed atoms construct two fixed walls to prevent the atoms escaping from the system. Adjacent to the fixed walls, the hot and cold reservoirs (heat source and heat sink) are established to create the temperature gradient. When the simulation structure is in an equilibrium state, the velocities of the atoms in the reservoir regions are adjusted artificially in order to introduce a constant heat flux into the normal simulation region, which is located between the heat source and heat sink. At each time step, the velocities of the particles in the heat source and heat sink are scaled by the same factor so that the net kinetic energy is added or reduced with the same amount from the two reservoirs. This way, a constant heat flux is introduced into the simulation domain and a temperature gradient can be established along the heat flow direction for the simulation time. From the heat flow and the temperature gradient, the thermal conductivity of the plumbene can be obtained from the Fourier law. The model is divided into numerous slabs with a length of 20-25 Å along the X axis direction to evaluate the local temperatures.

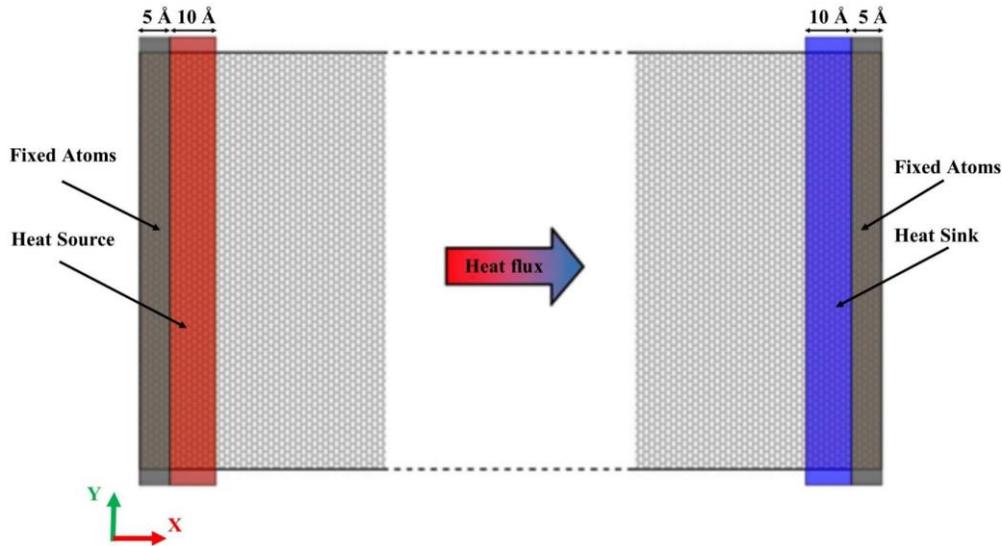

Figure 2: The simulation model for the NEMD calculations of the thermal conductivity.

A time step of 1 fs was used in all simulations to simulate the thermal conductivity of plumbene. The simulation box was first relaxed in the NVT (number of particles, volume, and temperature) ensemble for 100 ps. The heat source and heat sink were then controlled by Langevin thermostats at constant temperatures, and finally by the NVE (Microcanonical) ensemble to calculate the heat flux along the plumbene until the temperature profile becomes linear over 1 ns. To calculate the in-plane thermal conductivity of plumbene, the Fourier law was applied using Equation (11).

$$q = -k\frac{\partial T}{\partial x} \qquad (11)$$

where q is the heat flux obtained from the NEMD simulation, k is the thermal conductivity coefficient and ∂T/∂x is the temperature gradient.

To comprehensively assess the interaction between lead atoms, we employ three different potentials. Firstly, we utilize the optimized SW and Tersoff potentials, which have been specifically optimized for this study. These potentials have been carefully calibrated to accurately capture the behavior of plumbene. Additionally, we incorporate the EAM potential used by Das et al. [14] to explore its capability in reproducing the characteristic buckling hexagonal structure of plumbene. By employing these three distinct potentials, we aim to thoroughly analyze and understand the behavior of plumbene, allowing us to gain valuable insights into its structural properties.

In this study, we investigate the effects of temperature, strain, length, and width on the thermal conductivity of plumbene. To assess the impact of length on thermal conductivity, we vary the length from 300 Å to 1050 Å for a 300 Å wide sheet at 300 K. To study how width affects thermal conductivity, we use sheets with a fixed length of 300 Å and vary the width from 300 Å to 700 Å at room temperature. To examine the temperature dependence of thermal conductivity, we also computed the thermal conductivity with varying temperatures in the range of 300 K to 600 K while the size was kept fixed at 300 Å × 300 Å. Finally, to induce tensile strain, we move one fixed region at 0.5 Å/ps after equilibrium and expand the sample to the desired amount.

## 3. Evaluation of the potentials parameters optimization and the NEMD method

In order to check the accuracy of the NEMD simulation procedure for the thermal conductivity calculation of plumbene, the thermal conductivity simulation is performed according to the dimensions and operating conditions of Das et al. [14], which is the only published research to date on the estimation of plumbene's thermal conductivity. The results obtained from our MD simulations using EAM potential are compared with the published data in Figure 3. This figure clearly shows that simulation results are in good agreement.

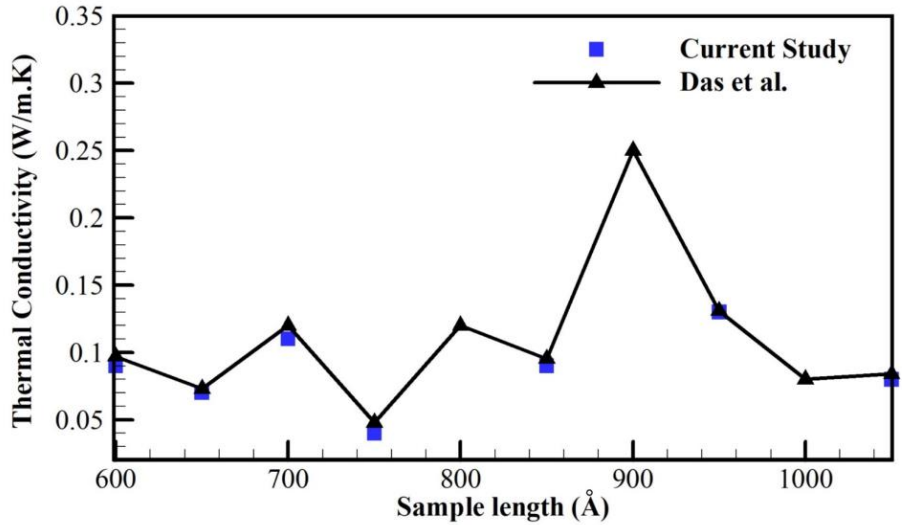

Figure 3: Comparisons of the thermal conductivity of plumbene sheets with 300 Å width obtained from NEMD simulations using EAM potential in the current study and Das et al. [14] at 300 K.

While the obtained results align with the findings presented in reference [14] when utilizing the EAM potential, our study goes further by conducting a detailed examination and comparison of the atomic structure of plumbene across all three potentials. We aim to provide a comprehensive analysis that delves deeper into the intricacies of plumbene's atomic arrangement.

MD simulation results show that after equilibrium, sheets simulated with SW and Tersoff potentials retain their honeycomb structure and buckling height. Contrastingly, the application of EAM potential yields a deviation, failing to preserve the characteristic hexagonal structure. In addition, the average buckling height, average bond length, and average lattice constant of LB plumbene were calculated using the results of *Ab initio* MD simulations and MD simulations. Calculated structural parameters for LB plumbene at room temperature are tabulated in Table 3. Our computed structural parameters with *Ab initio* MD simulation and MD simulations with the optimized SW and Tersoff potentials are consistent with the previous results [2,3,5,6,9]. However, equilibrium lattice constant, bond length and buckling height obtained from EAM potential are significantly different.

Table 3. Lattice constant, bond Length and buckling height for low-buckled plumbene at 300 K

|  | Lattice constant (Å) | Bond length (Å) | Buckling height (Å) |
|---|---|---|---|
| Current study, *Ab initio* MD simulation | 4.98 | 2.99 | 1.06 |
| Current study, MD simulation with EAM potential | 3.72 | 3.22 | 2.03 |
| Current study, MD simulation with optimized Tersoff potential | 4.81 | 2.96 | 1.07 |
| Current study, MD simulation with optimized SW potential | 4.93 | 2.99 | 0.99 |
| Zhao et al. (2016) [2] | 4.93 | 3.00 | 0.93 |
| Lu et al. (2016) [3] | 4.93 | - | - |
| Zhang et al. (2018) [5] | 4.93 | 3.00 | 0.99 |
| Li et al. (2019) [6] | 4.93 | - | 0.99 |
| Vivek et al. (2021) [9] | 5.08 | 3.09 | 0.95 |

The structural parameters of LB plumbene within the temperature range of 100 K to 500 K were also determined through *Ab initio* MD simulations. These calculated parameters were then compared with the results obtained from MD simulations, as depicted in Figure 4. The graphs illustrate that all structural parameters exhibit an increase with rising temperature. Moreover, the NEMD simulations conducted using the optimized SW and Tersoff potentials yield consistent results with the *Ab initio* MD simulations.

Notably, the SW potential demonstrates a closer correspondence to the *Ab initio* results, with maximum errors of only 0.4%, 1.08%, and 8% for the bond length, lattice constant, and buckling height, respectively. On the other hand, the Tersoff potential yields slightly larger discrepancies, with maximum errors of 1.5%, 3.8%, and 7% for the same parameters. In contrast, the results obtained with the EAM potential differ significantly across all temperatures.

The findings presented in Table 3 and Figure 4 affirm the consistency of the optimized SW and Tersoff potentials with the actual structure of LB plumbene. Furthermore, the optimized SW potential outperforms the optimized Tersoff potential in terms of accuracy. Considering the observed disparity between the actual LB plumbene structure and the simulated equilibrium structure using the EAM potential, the subsequent section simulates the thermal conductivity of plumbene using the Tersoff and SW potentials with the optimized parameters.

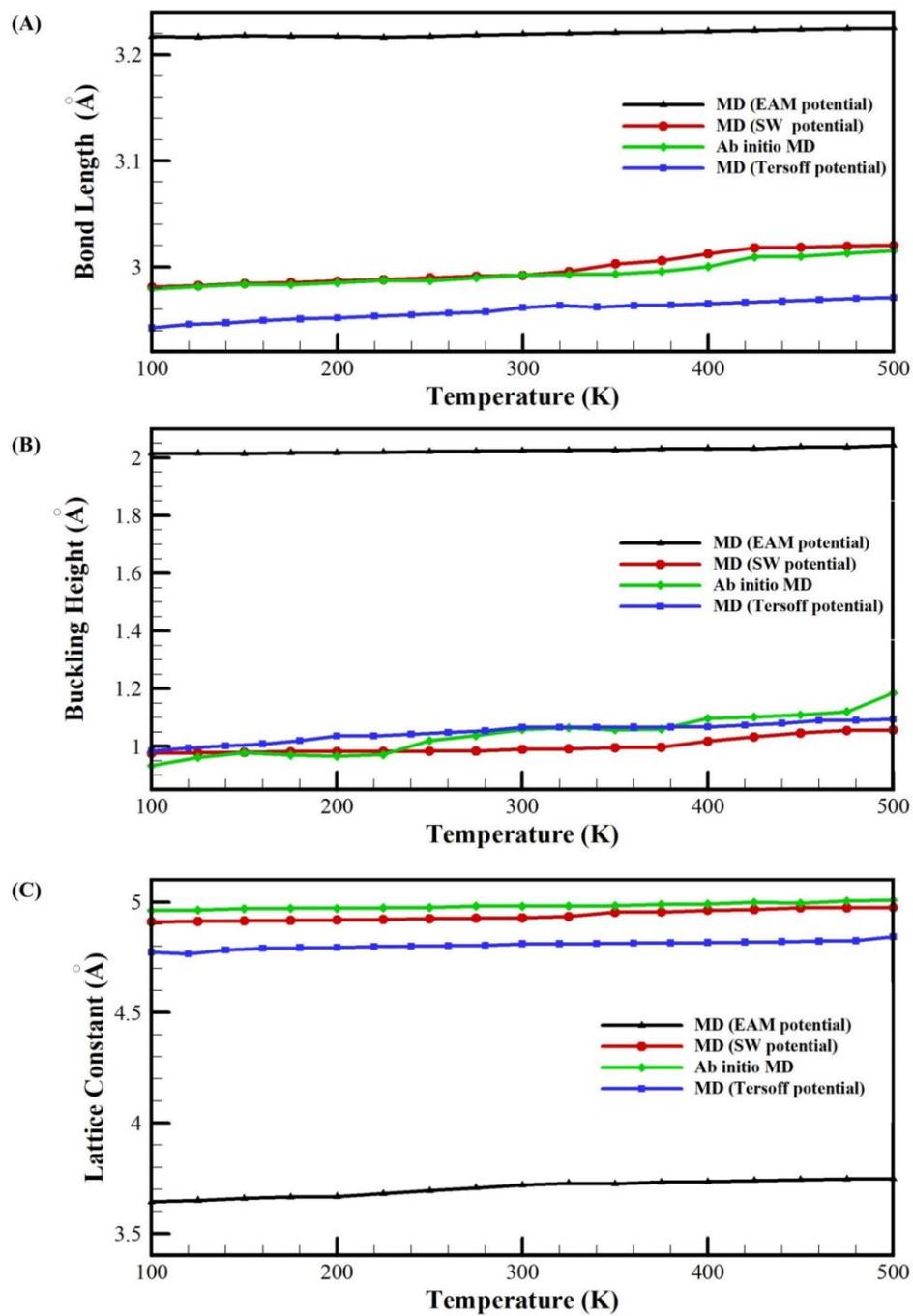

Figure 4: (A) The bond-length, (B) Buckling-height, and (C) Lattice-constant as a function of temperature from MD simulations with different potentials and *Ab initio* MD simulations.

## 4. Results and Discussion

The previous section's results clearly indicate that the available EAM potential for lead lacks the required accuracy and quality to faithfully reproduce the intricate structure of plumbene. Consequently, the present study focuses on investigating the thermal conductivity of plumbene using the Tersoff and SW potentials with the optimized parameters obtained earlier. Furthermore, we explore the influence of sample size, temperature, and strain on the thermal conductivity for different edge shapes of plumbene sheets.

Figure 5 illustrates the results for a plumbene sheet with a width of 300 Å and varying lengths at 300 K. Notably, the thermal conductivity values obtained from the SW potential tend to be higher than those from the Tersoff potential. However, since there are no experimental measurements available for plumbene's thermal conductivity, it is challenging to determine which set of outcomes is more accurate.

The observed thermal conductivity of plumbene falls within the range of 0.7 - 8 W/m.K at room temperature, which is considerably smaller than that of bulk lead (34.7 W/m.K). Similar reductions in thermal conductivity have been observed in other 2D materials, such as silicene compared to bulk silicon [33] and stanene compared to bulk tin [34]. This substantial reduction in thermal conductivity can be attributed to increased phonon-surface scattering in low-dimensional semiconducting nanostructures.

Figure 5 also highlights that, in general, the thermal conductivity of plumbene increases with sample length for both the Tersoff and SW potentials. Previous studies on other group IVA 2D materials have reported similar trends of thermal conductivity increasing with sample length [15-20, 33, 35]. In contrast, Figure 3 demonstrates fluctuating behavior when employing the EAM potential. These findings strongly suggest that the fluctuating behavior observed with the EAM potential is likely incorrect. Moreover, the optimized SW and Tersoff potentials, as demonstrated in the previous section, are more capable of accurately simulating the equilibrium structural parameters compared to the EAM potential. Thus, the EAM potential's inadequacy in predicting the equilibrium structural parameters of LB plumbene may be the underlying cause of its unusual fluctuating thermal conductivity with respect to sample length.

In addition, Figure 5 shows that plumbene sheets with zigzag edges exhibit higher thermal conductivity than those with armchair edges implying a universal edge-shape dependence of thermal conductivity in the 2D-materials sheets [36-38].

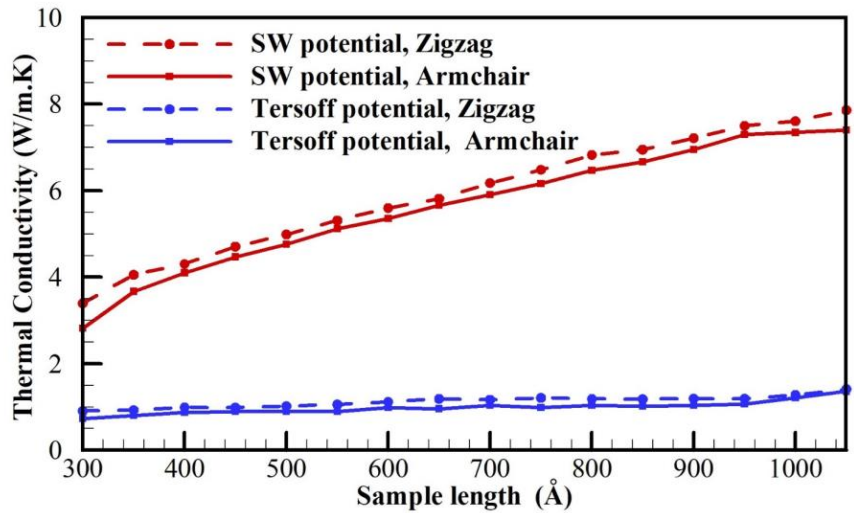

Figure 5: The thermal conductivities of plumbene sheets with a width of 300 Å as a function of length at 300 K.

Figure 6 displays the thermal conductivity of plumbene sheets with different widths. This result pertains to samples that are 300 Å in length at 300 K. The results from both potentials indicate that the width of the plumbene samples has no perceptible impact on their thermal conductivity. The observed effects can be attributed to periodic boundary conditions in the simulation. Previous NEMD simulations of other 2D group IVA monolayers have also shown that the thermal conductivity at 300 K becomes width independent [16, 39]. Moreover, the simulation results show again that plumbene sheets with zigzag edges possess higher thermal conductivity than those with armchair edges, as discussed above.

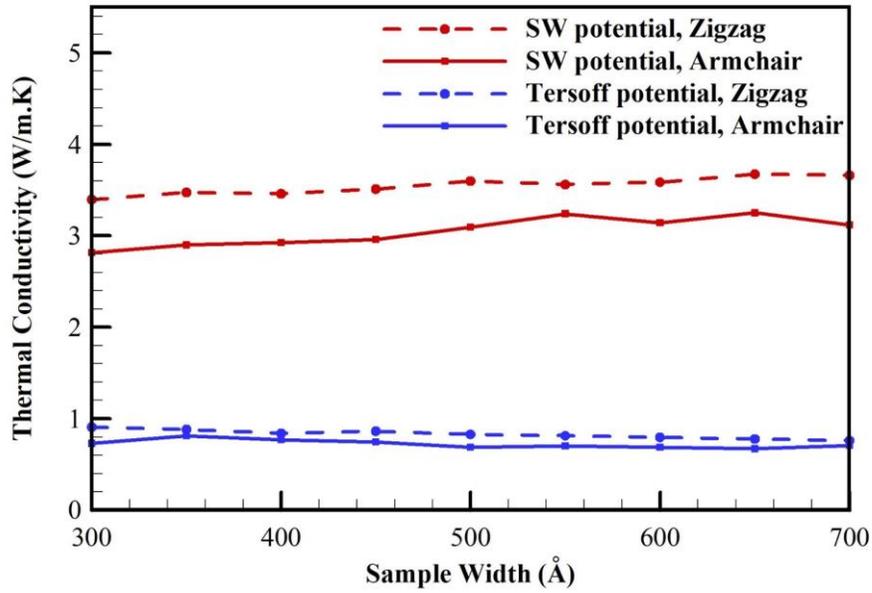

Figure 6: The thermal conductivities of plumbene sheets with a length of 300 Å as a function of width at 300 K.

Figure 7 depicts the variation of the thermal conductivity of plumbene with temperature for a 300 Å × 300 Å sample. This figure shows that increasing the temperature above 300 K, leads to a reduction in thermal conductivity for both armchair and zigzag configurations at almost similar rates. This is a common trend observed in other 2D materials [17, 20, 36]. This phenomenon can be attributed to the Umklapp phonon-phonon scatterings. The obtained results show that the thermal conductivity of the zigzag configuration, obtained from SW potential, decreases from 3.4 W/m.K to 2 W/m.K when the temperature increases from 300 K to 600 K. Moreover, as shown in Figure 7, zigzag configurations exhibit a higher thermal conductivity than the armchair configurations at all temperatures, which is in agreement with the behavior of other 2D nanostructures [36,38]

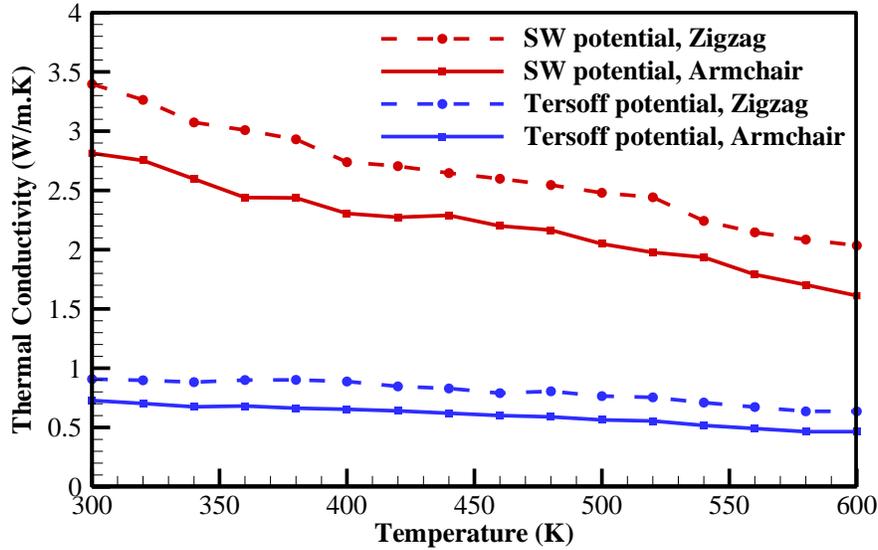

Figure 7: The thermal conductivities of a 300 Å × 300 Å plumbene sheet as a function of temperature.

Previous studies have demonstrated that strain plays a significant role in shaping the thermal conductivity of various 2D materials. The application of stress or strain offers a mechanism to finely tune the thermal conductivity of materials, as evidenced by these investigations [33, 40-44]. Notably, the response of different 2D materials to tensile stress can exhibit distinct behaviors. For instance, in graphene, the thermal conductivity consistently decreases with increasing tensile strain [40-43]. Conversely, in buckled silicene, an initial small increase in thermal conductivity is observed with the rise in tensile strain, followed by a subsequent decrease at larger strains [33, 44]. These findings emphasize the importance of strain in manipulating the thermal conductivity of 2D materials. By applying controlled tensile stress, it becomes possible to precisely modulate the thermal properties of these materials, allowing for tailored applications in various technological domains.

Here, we apply uniaxial tensile strain on the plumbene sheet of 300 Å ×300 Å at 300 K. Heat fluxes are applied in the strained direction. The calculated thermal conductivities in the strained direction obtained by both Tersoff and SW potentials are illustrated in Figure 8. This figure shows that the Tersoff and SW potentials present different tensile failure limits of 3.4% and 20%, respectively. In addition, tensile strain decreases thermal conductivity monotonically when Tersoff potential is used, while the results of SW potential show that initially the thermal conductivity increases with increasing strain from 0 to 5% and with further increasing strain, the thermal conductivity starts to decrease. Since the optimized SW potential is able to predict the structural parameters better (as discussed in Sec. 3), and the buckled structure of plumbene is more similar to silicene than the planar structure of graphene, we believe the thermal behavior under strain which is obtained by SW potential would be more accurate. Earlier reported NEMD simulations of other nanostructures in literature also demonstrate that the thermal conductivity of 2D materials shows notably different temperature-dependence with various potentials [45,46].

By applying the optimized SW potential, the maximum thermal conductivity appears at 5% strain and its value is about 12% (2%) larger than that for the unstrained armchair (zigzag) configuration. At a tensile strain of 0.2, the thermal conductivity drops more than 64% (46%) compared to the strain-free value for the armchair (zigzag) configuration. The thermal behavior of plumbene under strain obtained by SW potential could be attributed to its initial buckled configuration. At small tensile strains, the buckled configuration becomes less buckled. As the buckled structure flattens, the scattering channels for phonons, especially out-of-plane modes are reduced, resulting in an enhanced thermal conductivity. At large tensile strains, the Pb-Pb bonds are severely stretched. Increasing the Pb-Pb bond length will weaken the interatomic interactions in the in-plane direction. As a consequence, in-plane stiffness decreases and so does the thermal conductivity.

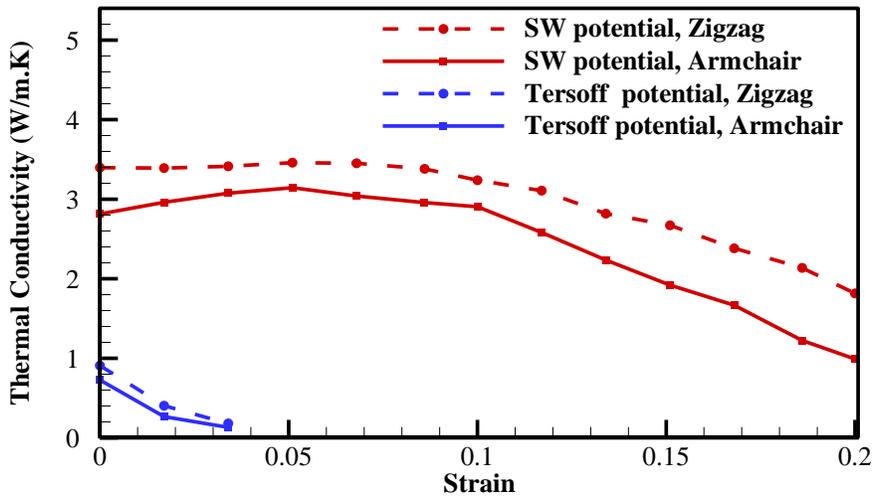

Figure 8: Effect of strain on thermal conductivity of a 300 Å × 300 Å plumbene sheet at 300 K.

## 5. Conclusion

In this study, we examined the thermal conductivity of low-buckled plumbene with different edge shapes by utilizing non-equilibrium molecular dynamics. The results revealed that no existing interatomic potential could precisely capture the unique features of plumbene. Consequently, we optimized the parameters of the Tersoff and Stillinger-Weber potentials using ANNs for a single-layer Pb sheet. This methodology enabled accurate reproduction of plumbene's low buckling structure as obtained from *Ab initio* MD calculations. Our calculated equilibrium buckling height of 0.99(1.07) Å and lattice constant of 4.93 (4.81) Å, derived from the optimized SW (Tersoff) potential, match reasonably well with the values obtained from previous DFT simulations. However, the optimized SW potential demonstrated a superior capacity to predict accurate structural results.

We then applied the optimized SW and Tersoff potentials to calculate the thermal conductivity of plumbene, exploring the relationship between plumbene's length, temperature, width, and strain with its thermal conductivity. The thermal conductivity of low-buckled plumbene calculated with the optimized SW potential was, on average, 4 times larger than the thermal conductivity obtained with the optimized Tersoff potential. The thermal conductivity of plumbene sheets (with widths of 300 Å and lengths ranging from 300 Å to 1050 Å) at room temperature ranged from 2.8 – 7.8 W/m.K with the optimized SW potential and from 0.7 – 1.4 W/m.K with the optimized Tersoff potential. Regardless of the results' magnitude, both the optimized SW and Tersoff potentials affirmed that an increase in the length of plumbene leads to an increase in its thermal conductivity.

Due to periodic boundary conditions, plumbene's thermal conductivity did not exhibit a clear trend corresponding to changes in sheet width. Furthermore, we observed that an increase in temperature led to a decrease in plumbene's thermal conductivity. As the temperature increased from 300 K to 600 K, the thermal conductivity of a 300 Å × 300 Å sample with a zigzag configuration decreased from 3.4 W/m.K to 2 W/m.K. Consistently, plumbene sheets with zigzag edges demonstrated higher thermal conductivity than those with armchair edges, aligning with the behavior of other 2D nanostructures.

We also examined the effect of applying uniaxial tensile strain on plumbene. When we employed the Tersoff potential, tensile strain resulted in a monotonous decrease in thermal conductivity, similar to the behavior of graphene under tensile strain. On the other hand, the results from the SW potential showed that thermal conductivity initially increased with strain from 0 to 5% but started to decrease with further strain, a behavior akin to silicene under tensile strain. Given the optimized SW potential's better prediction of plumbene's structural parameters and the closer resemblance of plumbene's buckled structure to silicene than to graphene's planar structure, we posit that the results obtained by the optimized SW potential in this research are more reliable.

## 6. Conflicts of interest

There are no conflicts of interest to declare.